\documentclass[sigconf]{acmart}

\setcopyright{none}
\renewcommand\footnotetextcopyrightpermission[1]{}

\usepackage{tabularx}
\usepackage{multirow}
\usepackage{arydshln}

\definecolor{hintcolor}{rgb}{0, 0, 0.7} 
\definecolor{answercolor}{rgb}{0.7, 0, 0} 

\AtBeginDocument{%
  }

\begin{document}

\title{Aligned Query Expansion: Efficient Query Expansion for Information Retrieval through LLM Alignment}

\author{Adam Yang}
\authornote{Work done during internship at Spotify.}
\affiliation{%
  \institution{Mistral AI}
  \country{UK}
}
\email{adamxinyuyang@gmail.com}

\author{Gustavo Penha}
\affiliation{%
  \institution{Spotify}
  \country{Netherlands}}
\email{gustavop@spotify.com}

\author{Enrico Palumbo}
\affiliation{%
  \institution{Spotify}
  \country{Italy}
}
\email{enricop@spotify.com}

\author{Hugues Bouchard}
\affiliation{%
  \institution{Spotify}
  \country{Spain}
}
\email{hb@spotify.com}







\begin{abstract}
With the breakthroughs in large language models (LLMs), query generation techniques that expand documents and queries with related terms are becoming increasingly popular in the information retrieval field. Such techniques have been shown to improve the effectiveness of traditional lexical retrieval methods by dealing with the vocabulary mismatch problem. Recent work has found that generating queries with a greedy decoding strategy can produce sub-optimal queries, including hallucinations, and proposed to filter out queries before expansion. This `generate-then-filter' approach is costly, as it requires generating multiple queries and applying a relevance model to all of them and does not teach the LLM which of the generated queries is more effective for expansion. To overcome such limitations, we propose Aligned Query Expansion (AQE), a novel approach to enhance query expansion for passage retrieval in open-domain question answering. AQE leverages recent techniques in LLM alignment to fine-tune models for generating query expansions that directly optimize the effectiveness of the retrieval task, eliminating the need for additional filtering steps. This alignment ensures that queries are more relevant, reducing computational costs while improving retrieval effectiveness. Empirical evaluations show that AQE outperforms baseline models for query expansion in both in-domain and out-of-domain settings, demonstrating significant improvements in retrieval effectiveness.
\end{abstract}





\maketitle
\pagestyle{plain}

\section{Introduction}

Information retrieval (IR) systems play a pivotal role in organizing and accessing the vast amounts of information available today. From search engines to digital libraries, the effectiveness of these systems hinges on their ability to accurately match user queries with relevant documents. However, one of the longstanding challenges in IR is the \emph{vocabulary mismatch problem}, where the terms used in user queries do not directly align with those in relevant documents. This discrepancy can lead to decreased retrieval performance, as traditional lexical matching techniques may fail to identify pertinent information.

To address this issue, query expansion techniques have been employed to enhance IR systems by enriching user queries with additional related terms. Traditionally, these methods relied on statistical measures such as term co-occurrence~\cite{xu2000improving} or leveraged external resources like thesauri to introduce synonyms and related concepts. While these approaches can mitigate the vocabulary mismatch to some extent, they often suffer from \emph{over-expansion}, where the addition of irrelevant terms dilutes the query's focus and negatively impacts retrieval accuracy.

The advent of large language models (LLMs) has revolutionized query expansion by enabling more sophisticated and contextually relevant term generation. For instance, models like T5~\cite{raffel2019exploring} have been used by Doc2Query~\cite{nogueira2019doc2query}, which generates queries to augment document representations before indexing. Similarly, LLMs have been employed to expand user queries themselves, producing enriched versions that capture a broader semantic scope~\cite{wang2023query2doc}. These generative approaches have demonstrated significant improvements in retrieval effectiveness by addressing the vocabulary mismatch problem more robustly than traditional methods.

However, leveraging LLMs for query expansion introduces new challenges. A notable issue is \emph{hallucination}, where the model generates plausible but factually incorrect or irrelevant terms with confidence \citep{huang2023survey,yang2023bayesian,yang2024just}. This can degrade retrieval performance by introducing noise into the expanded queries. To combat this, recent methodologies adopt a \textit{generate-then-filter} paradigm. Techniques like Doc2Query-{-}~\cite{gospodinov2023doc2query} and Expand and Rerank (EAR)~\cite{chuang2023expand} generate multiple query candidates and subsequently apply a reranking or filtering step to retain only the most relevant expansions. While effective, this approach incurs significant computational overhead, as it necessitates generating numerous queries and evaluating each with a relevance model. Moreover, it does not inherently guide the LLM to prioritize query expansions that are most beneficial for the retrieval task, leading to inefficiencies and increased latency—particularly problematic in real-time IR systems with stringent performance requirements.

Beyond the computational costs, the generate-then-filter strategy also faces limitations in adaptability. Models are constrained by their inability to learn and prioritize the generation of highly effective query expansions based on retrieval outcomes. This results in a scenario where the system may not consistently produce expansions that align optimally with downstream retrieval objectives, thereby limiting overall performance gains.

Given these challenges, there is a compelling need for a more streamlined and effective approach to query expansion that leverages the strengths of LLMs while mitigating their drawbacks. This motivates the central question of our research: \emph{Can we directly optimize synthetic query generation for downstream retrieval effectiveness without relying on costly filtering steps?}

In this paper, we introduce \textbf{Aligned Query Expansion (AQE)}, a novel methodology designed to enhance query expansion for passage retrieval in open-domain question-answering systems. AQE harnesses recent advancements in LLM alignment to fine-tune language models specifically for generating query expansions that are inherently more effective for retrieval tasks. By aligning the LLM's generation process with the ultimate goal of retrieval effectiveness, AQE eliminates the need for subsequent filtering steps, thereby reducing computational costs and latency.

Our approach diverges from traditional generate-then-filter methods by directly optimizing the query generation process to produce expansions that are relevant and diverse, tailored to enhance retrieval performance. This alignment ensures that the generated queries contribute positively to the retrieval task, mitigating the risks of hallucination and over-expansion of generative models.

Empirical evaluations conducted across various datasets demonstrate that AQE not only surpasses baseline models in retrieval effectiveness both in-domain and out-of-domain but also achieves significant reductions in computational overhead. Specifically, AQE reduces latency by approximately 70\% compared to existing approaches, underscoring its potential for deployment in real-time IR systems. In fact, concurrent work has deployed aligned query generation methods in large scale industrial settings leading to significant gains in online metrics~\cite{peng2024large}.

In summary, our contributions are threefold:
\begin{itemize}
    \item We introduce \textbf{Aligned Query Expansion (AQE)}, a novel query expansion framework that leverages LLM alignment techniques to generate more effective query expansions without the need for extensive filtering.
    \item We provide comprehensive empirical evidence showcasing AQE's superior performance in both in-domain and out-of-domain retrieval tasks, highlighting its robustness and generalizability across diverse datasets.
    \item We demonstrate that AQE significantly reduces computational memory and time costs, offering a more efficient and scalable solution for real-world systems.
\end{itemize}

Through AQE, we aim to advance the field of information retrieval by providing a more efficient and effective means of query expansion, paving the way for faster and more accurate retrieval systems in the era of large language models.

\section{Related Work}
In this section, we provide relevant references for related work on document expansion and query expansion, with a particular focus on generative approaches. Then, we introduce LLM alignment, highlighting the methodologies that are applied in this paper. 

\subsection{Document Expansion with Query Generation}
Sparse retrieval systems that match keywords between the user's query and a document are highly cost effective, but suffer from the vocabulary mismatch problem. Document expansion techniques~\cite{efron2012improving,singhal1999document} are often used to improve the retrieval effectiveness of such systems, by adding terms that are related to the document and increases the chance of matching relevant queries (i.e. by adding synonyms, closely related concepts, etc...).
\paragraph{Doc2Query~\cite{nogueira2019doc2query}} introduces the idea of fine-tuning a seq2seq model to generate synthetic queries that can be used for document expansion. The authors show that the approach is highly effective, significantly improving both sparse and dense retrieval models. Similar approaches have shown to be able to reduce retrievability bias as well~\cite{penha2023improving}. In general, though, generative models are prone to hallucinations, which can result in irrelevant queries and document expansion that can potentially damage the retrieval effectiveness.
\paragraph{Doc2Query-{-}~\cite{gospodinov2023doc2query}} builds on top of Doc2Query with the aim of pruning potentially irrelevant synthetically generated queries. This is accomplished by training an additional model that learn query-document relevant and is used to filter out poor quality synthetic queries before the document augmentation. The authors show that this improves retrieval effectiveness and decreases the index size. 
However, this approach requires to run an additional re-ranking step for each document in the index, which can be expensive for real-world scenarios where corpora include billions of documents.

\subsection{Query Expansion}
An alternative solution to tackle the vocabulary mismatch problem is to expand the user's query with related terms at retrieval time. While traditional solutions would typically rely on analyzing documents to find frequently co-occurring terms and related keywords~\cite{xu2000improving}, recent approaches leverage the power of generative models to produce query expansions.

\paragraph{Generation-Augmented Retrieval (GAR)}
This approach introduced by \citet{mao2020generation} enhances query formulation by generating additional contexts through a pre-trained language model (PLM). Unlike traditional query reformulation methods that require external resources or downstream feedback, GAR augments queries with relevant contexts, such as answers, sentences containing the answers, and passage titles. These generated contexts are appended to the original query, expanding its semantic scope for more effective retrieval. GAR is trained using standard sequence-to-sequence learning, with the question as input and relevant contexts (answers, sentences, or titles) as outputs. During retrieval, queries appended with these generated contexts are independently processed, and their results are fused. Simple fusion strategies, such as combining top-retrieved passages from different context-augmented queries, consistently improve retrieval effectiveness.

\paragraph{Expand and Rerank (EAR)}
\citet{chuang2023expand} proposed \textit{Expand and Rerank (EAR)}, a method that trains a query reranker to select query expansions that lead to better retrieval results. Specifically, the reranker $\mathcal{M}(q,e)$ is trained to predict the relative ranking of expansions $e$ by learning from the rank differences between pairs of expansions $(e_i, e_j)$. The objective function for training the reranker is formulated as:
\begin{align}
    \mathcal{L}_\text{Rank} = \sum_{i,j\in[1,n], r_i < r_j} \max(0, \mathcal{M}(q,e_i) - \mathcal{M}(q,e_j) + \alpha(r_j-r_i)),
\end{align}
where $r_i$ and $r_j$ represent the true document retrieval ranks of expansions $e_i$ and $e_j$, respectively, and $\alpha$ is a scaling factor. They also experimented with a reranker that incorporates the top-1 retrieved passage as an additional input to improve the ranking process.

EAR demonstrated strong performance when combined with the \textit{GAR generator}. However, a key drawback is its computational cost, as it requires the generation of many query expansions (e.g., 50) followed by reranking to filter the best queries. This makes the filtering approach less efficient for real-time retrieval tasks where computational resources and latency are critical factors.

\subsection{LLM Alignment}

As large language models (LLMs) continue to advance, ensuring their outputs align with human preferences or other predefined reward functions has become paramount. LLM alignment involves adjusting these models to produce responses that are not only accurate but also consistent with human expectations \citep{ouyang2022training,touvron2023llama,touvron2023llama2,dubey2024llama,shi2024instruction}. This alignment is crucial for applications ranging from conversational agents to information retrieval systems, where the quality and appropriateness of the generated content directly impact user experience. In this section, we delve into various strategies for aligning LLMs with human preferences, including reward modeling, Best-of-$n$ (BoN) sampling, Reinforcement Learning from Human Feedback (RLHF), and Direct Preference Optimization (DPO).

\paragraph{Reward modeling }
In LLM alignment, we typically model human preference using a reward model \citep{ouyang2022training}. Specifically, for a pair of responses to a prompt $(x,y_w)$ and $(x,y_l)$, we define the human preference model (the Bradley-Terry model) as
\begin{align} \label{eq:preference}
    P(y_w > y_l) = \frac{e^{r_\theta(x,y_w)}}{e^{r_\theta(x,y_w)} + e^{r_\theta(x,y_l)}} = \sigma(r_\theta(x,y_w) - r_\theta(x,y_l)),
\end{align}
where $r_\theta$ is the reward model and $\sigma(\cdot)$ is the sigmoid function. Then we simply perform maximum log-likelihood optimization to learn the reward model given a fixed preference dataset
\begin{align}
    \max_\theta \mathbb{E}_{x,y_w,y_l} [\log\sigma(r_\theta(x,y_w) - r_\theta(x,y_l))].
\end{align}

\paragraph{Best-of-$n$ (BoN) sampling}
BoN sampling \citep{stiennon2020learning,ouyang2022training,coste2023reward,yang2024bayesian} is a decoding strategy to align LLM outputs with a given reward model without further fine-tuning the LLM policy. For any test prompt, BoN samples $n$ responses, and uses the reward model to rank the responses and select the \textit{best} one, which has the highest reward. 
The filtering approach used in query expansion methods, such as EAR and Doc2Query-{-}, operate similarly to BoN sampling in that both methods generate multiple candidates and then apply a ranking mechanism to filter out less relevant responses.

\paragraph{Reinforcement Learning from Human Feedback (RLHF) }
In RLHF \citep{ouyang2022training,touvron2023llama,touvron2023llama2}, the goal is to fine-tune the LLM policy, $\pi_\phi(y \mid x)$, to maximize the expected proxy reward, incorporating a regularization term to maintain behavioral closeness to a reference policy. The optimization problem is formulated as
\begin{align}
    \max_{\phi} \mathbb{E}_{x \sim \mathcal{D}, y \sim \pi_\phi(y \mid x)} [r_\theta(x, y)] - \beta \text{KL}(\pi_\phi(y \mid x) \parallel \pi_{\text{ref}}(y \mid x)),
\end{align}
where $\pi_{\text{ref}}(y \mid x)$ represents the reference policy before the application of RLHF. 

\begin{figure*}[ht]
    \centering
    \includegraphics[width=\textwidth]{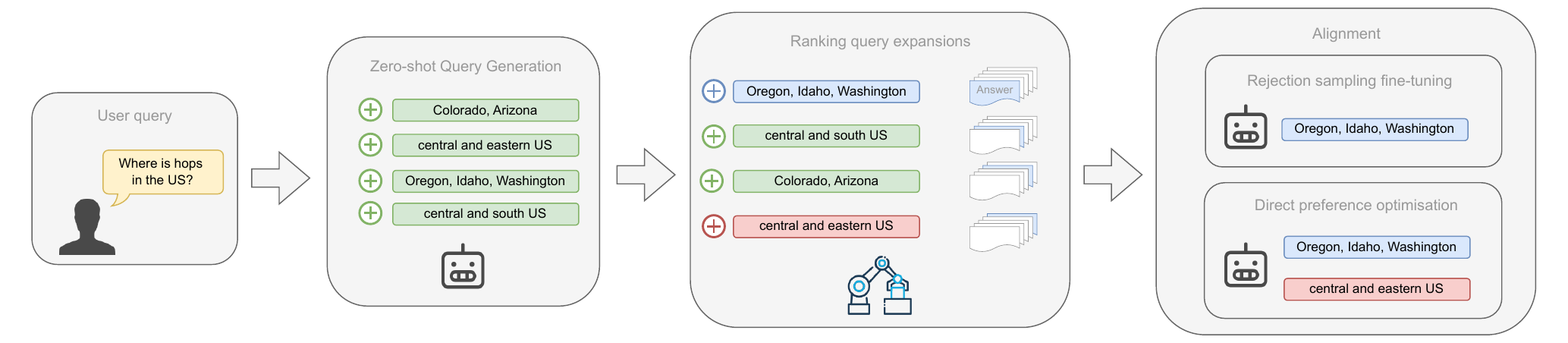}
    \label{fig:bon_single_var}
    \caption{Aligned Query Expansion (AQE) training pipeline. The pipeline begins with the generation of several query expansions using zero-shot prompting of a large language model, followed by ranking the expansions based on their retrieval effectiveness when used in a retrieval model. The top and bottom-ranked expansions are then used to fine-tune the LLM via two alignment strategies: Rejection Sampling Fine-Tuning (RSFT) and Direct Preference Optimization (DPO).}
    \label{fig:pipeline}
\end{figure*}

\paragraph{Direct Preference Optimization} 
Reinforcement Learning from Human Feedback (RLHF) \citep{ouyang2022training} typically involves two stages: training a reward model based on human preferences and then fine-tuning the language model via reinforcement learning to maximize this reward signal while maintaining distributional similarity with the pre-trained model. Despite its successes, RLHF can be computationally expensive, unstable, and complex due to the requirement of learning a reward model and performing reinforcement learning simultaneously.

To address these challenges, \citet{rafailov2024direct} introduced Direct Preference Optimization (DPO) by directly optimizing the model to align with human preferences using a contrastive objective. The core insight behind DPO is its ability to implicitly model the reward signal directly within the policy's log-probabilities, eliminating the need for explicit reinforcement learning. This is achieved by reparameterizing the reward model and deriving an update rule that maximizes the log-probability of preferred responses while minimizing the likelihood of dispreferred ones. 


\section{Aligned Query Expansion}

In this section, we detail the training pipeline for \textbf{Aligned Query Expansion (AQE)}, as illustrated in Figure~\ref{fig:pipeline}. AQE consists of three main components: zero-shot query expansion generation, ranking of query expansions, and alignment through fine-tuning. 

\subsection{Zero-shot Query Expansion Generation}
Given a training dataset $\mathcal{D} = \{(q_i, d_i)\}_{i=1}^N$, where $q_i$ represents a user query and $d_i$ is the corresponding relevant document, our first step is to generate a diverse set of query expansions for each query $q_i$. We employ a zero-shot prompting strategy using a large language model (LLM) $M$ to ensure diversity in the generated expansions. For each query $q_i$, we generate $n$ query expansions $\{e_{i,1}, e_{i,2}, \dots, e_{i,n}\}$ by prompting the LLM with the prompt `To answer this query, we need to know:' together with the original query in front, where $n=50$ in our experiments. This approach leverages the LLM's generative capabilities to produce a wide range of semantically related expansions, enhancing the potential for retrieving relevant documents.

\subsection{Ranking Query Expansions}
After generating a set of query expansions, the next step is to evaluate their effectiveness in retrieving the relevant document $d_i$ for each query $q_i$. We retrieve a ranked list of documents using a sparse retrieval algorithm (BM25) and then evaluate each expansion based on the rank of the true document $d_i$ in this list.

For each query expansion $e_{i,j}$, we determine the rank of the true document $d_i$ in the list of top-retrieved documents. Let $\text{Rank}(e_{i,j})$ denote the rank of the true document when the retrieval is performed with expansion $e_{i,j}$. The higher the rank, the more effective the expansion. We then identify the best and worst expansions:
\begin{align}
    e_{i}^{\text{best}} &= \arg\min_{j} \ \text{Rank}(e_{i,j}), \\
    e_{i}^{\text{worst}} &= \arg\max_{j} \ \text{Rank}(e_{i,j}),
\end{align}
where $e_{i}^{\text{best}}$ is the expansion that results in the highest rank (i.e., lowest rank value) for the true document, and $e_{i}^{\text{worst}}$ is the expansion that results in the lowest rank (i.e., highest rank value).

By using the rank of the true document as the evaluation criterion, we can directly assess which expansions are most helpful for improving retrieval effectiveness.

\subsection{Alignment}
The alignment step aims to fine-tune the LLM to generate more effective query expansions by leveraging the ranked expansions. We explore two alignment strategies: \textbf{Rejection Sampling Fine-Tuning (RSFT)} and \textbf{Direct Preference Optimization (DPO)}.

\subsubsection{Rejection Sampling Fine-Tuning (RSFT)}
In the RSFT approach \citep{dubey2024llama}, we fine-tune the LLM $M$ using only the best-performing expansions. Let $\mathcal{E}_{\text{RSFT}} = \{e_{i}^{\text{best}}\}_{i=1}^N$ denote the set of top-ranked expansions across the training dataset. The objective is to maximize the likelihood of these high-quality expansions:
\begin{align}
    \mathcal{L}_{\text{RSFT}} = \sum_{i=1}^N \log P_{\phi}(e_{i}^{\text{best}} \mid q_i),
\end{align}
where $P_{\phi}(e \mid q)$ represents the probability of generating expansion $e$ given query $q$ under the LLM parameterized by $\phi$. The fine-tuning process involves optimizing $\phi$ to maximize $\mathcal{L}_{\text{RSFT}}$:
\begin{align}
    \phi^* = \arg\max_{\phi} \mathcal{L}_{\text{RSFT}}.
\end{align}

This selective fine-tuning encourages the model to preferentially generate expansions that have demonstrated high retrieval effectiveness.

\subsubsection{Direct Preference Optimization (DPO)}
The DPO approach \citep{rafailov2024direct} directly optimizes the LLM to prefer better expansions over worse ones using a contrastive objective. For each query $q_i$, we consider a pair of expansions $(e_{i}^{\text{best}}, e_{i}^{\text{worst}})$. The DPO objective encourages the model to assign higher probabilities to the best expansions compared to the worst ones.

The DPO loss for a single query-expansion pair is defined as:
\begin{align}
    & \mathcal{L}_{\text{DPO}}(q_i, e_{i}^{\text{best}}, e_{i}^{\text{worst}}) \\
    &= -\log \sigma\left(\beta \left(\log \frac{P_{\phi}(e_{i}^{\text{best}} \mid q_i)}{P_{\text{ref}}(e_{i}^{\text{best}} \mid q_i)}\right) - \beta \left(\log \frac{P_{\phi}(e_{i}^{\text{worst}} \mid q_i)}{P_{\text{ref}}(e_{i}^{\text{worst}} \mid q_i)}\right)\right),
\end{align}
where $\sigma(\cdot)$ is the sigmoid function, $P_{\text{ref}}(e \mid q)$ denotes the probability under the reference (pre-aligned) policy, and $\beta$ is a scaling hyperparameter controlling the strength of the preference.

The overall DPO loss across the entire training dataset is then:
\begin{align}
    \mathcal{L}_{\text{DPO}} = \sum_{i=1}^N \mathcal{L}_{\text{DPO}}(q_i, e_{i}^{\text{best}}, e_{i}^{\text{worst}}).
\end{align}

We optimize the LLM parameters $\phi$ to minimize $\mathcal{L}_{\text{DPO}}$:
\begin{align}
    \phi^* = \arg\min_{\phi} \mathcal{L}_{\text{DPO}}.
\end{align}

This objective ensures that the model learns to generate expansions that are more likely to be ranked higher in terms of retrieval effectiveness.

\subsection{Inference}
After the alignment stage, at inference time, AQE performs a single-shot greedy decoding using the original query. This approach allows for fast and efficient generation of high-quality query expansions, maintaining the speed benefits of zero-shot prompting while leveraging the alignment improvements obtained through fine-tuning.
By using greedy decoding, we avoid the need for multiple sampling passes as in the filtering approach, which significantly reduces computation time while still providing expansions that have been optimized to improve retrieval performance.

\section{Experimental Setup}
In this section, we go over the datasets, implementation details, and the evaluation procedure used in our experiments.

\subsection{Datasets}
Our experiments are conducted on four public datasets to ensure reproducibility.

\paragraph{Natural Questions} The Natural Questions benchmark \citep{kwiatkowski2019natural} contains over 300,000 real queries to Google, with annotations for the relevant Wikipedia page, long answer, and short answer. The questions seek factual information, and the Wikipedia page may or may not contain the information required to answer the question. The long answer is a bounding box on the Wikipedia page containing all information required to infer the answer, and the short answer is one or more entities that give a short answer to the question, or a boolean yes or no. 

\paragraph{TriviaQA} The TriviaQA benchmark \citep{joshi2017triviaqa} is a large-scale reading comprehension dataset containing over 650K question-answer-evidence triples, featuring complex, compositional questions that require cross-sentence reasoning to answer. The dataset is unique in that the questions are naturally occurring, well-formed, and collected independently of the evidence documents, which include both Wikipedia articles and web pages.

\paragraph{WebQA} WebQA \citep{chang2022webqa} is a benchmark for open-domain, multi-hop visual question answering that requires models to reason across both visual and textual modalities. The dataset mirrors how humans use the web to answer questions - by asking a question, choosing relevant sources, and generating a fluent language response.

\paragraph{Entity Questions} Entity Questions \citep{sciavolino2021simple} consists of simple, entity-centric questions based on facts from Wikidata. 4 The dataset was created by selecting 24 common relations from Wikidata, converting fact triples into natural language questions using manually defined templates, and sampling triples from the T-REx dataset where the triples are aligned with evidence sentences in Wikipedia.

\subsection{Implementation Details}
For practical implementation, we adopt the following settings:
\begin{itemize}
    \item \textbf{Language Model}: We use the T0 3B model~\cite{sanh2021multitask} for generating query expansions and further fine-tuning. We use this language model to ensure comparable results with previous work.
    \item \textbf{Generation Parameters}: Query expansions are generated using a zero-shot prompting strategy with temperature $\tau = 1.0$ and top-$k$ sampling with $k=50$ to ensure diversity.
    \item \textbf{Baselines}: We consider three baselines: (1) original query without query expansion; (2) zero-shot query expansion by prompting the LLM; (3) generate 50 query expansion and use a scoring model to filter the best. For filtering, we follow the setup in \citet{chuang2023expand} to fine-tune a DeBERTa V3 base \citep{he2020deberta} as the reranker. 
    \item \textbf{Alignment Hyperparameters}: Both RSFT and DPO are implemented using gradient-based optimization with AdamW optimizer~\cite{loshchilov2017decoupled}, learning rate $\eta = 5 \times 10^{-5}$, batch size $B = 16$, and fine-tuned for one epoch. For DPO, we set $\beta = 0.1$ to balance the influence of the reference policy.
\end{itemize}




\subsection{Evaluation Procedure}

\subsubsection{Evaluation Metric}
We use top-N retrieval accuracy metric on test sets for evaluation as in \citet{chuang2023expand}. Top-N accuracy is a common metric used to evaluate the performance of information retrieval systems, including recommendation engines and search systems. It measures the proportion of cases where the relevant document is found within the top N results returned by the system. Formally, it is defined as:
\begin{align*}
&\frac{\text{Number of queries where the relevant document is in top-N}}{\text{Total number of queries}}
\end{align*}
This metric is particularly useful in scenarios where users are expected to consider multiple top-ranked results. For instance, in recommendation systems, Top-5 or Top-10 accuracy measures how often the relevant document is presented within the top 5 or 10 suggestions. Top-N accuracy is an effective metric when the exact ranking of the result is less important, as long as the relevant document is present within the top few returned results.


\subsubsection{In-Distribution and Out-of-distribution}
For in-distribution experiments, we fine-tune the scoring model and perform alignment using the training split of a given dataset. We then evaluate the model’s performance on the test split of the same dataset to assess its effectiveness within the distribution it was trained on.

For out-of-distribution experiments, we assess the generalizability of the model by conducting evaluations on the test split of a different dataset than the one used for training. This allows us to measure how well each method adapts to unseen data that falls outside the training distribution.

\begin{table*}[ht]
\caption{Top-N retrieval accuracy comparison of different query expansion methods on the Natural Questions dataset. Bold indicates the highest effectiveness and superscripts denote statistically significant improvements over the corresponding model using Student's t-tests with Bonferroni correction.}
    \label{tab:natural}
    \centering
    \begin{tabular}{lllllllll}
        \toprule
        & & Model & Top-1 & Top-5 & Top-10 & Top-20 & Top-50 & Top-100 \\
        \midrule
        \multirow{3}{*}{Baselines} & \footnotesize{(0)}&Original query & 22.1 & 43.8 & 54.5 & 62.9$^{1}$ & 72.7 & 78.3 \\
        & \footnotesize{(1)}&Query expansion (zero-shot) & 23.5 & 44.5 & 54.0 & 61.4 & 71.0 & 77.5 \\
        & \footnotesize{(2)}&Filtering & 28.5$^{013}$ & 51.4$^{013}$ & 59.9$^{013}$ & 67.1$^{013}$ & 76.0$^{013}$ & 81.0$^{013}$ \\ 
        \hdashline \rule{0pt}{2.3ex}
        \multirow{3}{*}{AQE (Ours)} & \footnotesize{(3)}&RSFT  & 25.2$^{01}$ & 47.9$^{01}$ & 57.0$^{01}$ & 65.1$^{01}$ & 73.9$^{1}$ & 78.9$^{1}$ \\
        & \footnotesize{(4)}&DPO  & 29.8$^{013}$ & 51.9$^{013}$ & 60.6$^{013}$ & 68.7$^{013}$ & 76.8$^{013}$ & 81.1$^{013}$ \\
       &  \footnotesize{(5)}&RSFT + DPO  & \textbf{30.8$^{0123}$} & \textbf{53.6$^{0123}$} & \textbf{62.7$^{01234}$} & \textbf{70.0$^{0123}$} & \textbf{77.7$^{0123}$} & \textbf{82.4$^{01234}$} \\
        \bottomrule
    \end{tabular}
    \vspace{10pt}
\end{table*}

\begin{table*}[ht]
    \caption{Top-N retrieval accuracy comparison of different query expansion methods on the TriviaQA dataset. Bold indicates the highest effectiveness and superscripts denote statistically significant improvements over the corresponding model using Student's t-tests with Bonferroni correction.}
    \label{tab:trivia}
    \centering
    \begin{tabular}{lllllllll}
        \toprule
        & & Model & Top-1 & Top-5 & Top-10 & Top-20 & Top-50 & Top-100 \\
        \midrule
        \multirow{3}{*}{Baselines} & \footnotesize{(0)}& Original query & 46.3$^{2}$ & 66.3$^{123}$ & 71.7$^{123}$ & 76.4$^{123}$ & 80.6$^{12}$ & 83.2$^{12}$ \\
        & \footnotesize{(1)}&Query expansion (zero-shot) & 46.3$^{2}$ & 63.8$^{2}$ & 69.7$^{2}$ & 74.5$^{2}$ & 79.4 & 82.5 \\
        & \footnotesize{(2)}&Filtering & 42.6 & 62.1 & 68.1 & 73.7 & 79.0 & 82.3 \\  

        \hdashline \rule{0pt}{2.3ex}
        
        \multirow{3}{*}{AQE (Ours)} & \footnotesize{(3)}&RSFT  & 47.6$^{012}$ & 65.0$^{12}$ & 70.5$^{12}$ & 75.4$^{12}$ & 80.3$^{12}$ & 83.1$^{12}$ \\
        & \footnotesize{(4)}&DPO & \textbf{50.2}$^{0123}$ & \textbf{67.9}$^{0123}$ & \textbf{72.9}$^{0123}$ & \textbf{77.0}$^{123}$ & {81.2}$^{0123}$ & \textbf{83.7}$^{0123}$ \\
        & \footnotesize{(5)} & RSFT + DPO & {49.8}$^{0123}$ & {67.5}$^{0123}$ & {72.7}$^{0123}$ & \textbf{77.0}$^{123}$ & \textbf{81.3}$^{0123}$ & \textbf{83.7}$^{0123}$ \\
        \bottomrule
    \end{tabular}
    \vspace{10pt}

\end{table*}

\section{Results}
In this section, we report the main results of our experimentation, showing the effectiveness and the efficiency of the proposed solution compared to alternative approaches.

\subsection{In-distribution Training and Evaluation}
We first evaluate our approach by training and testing within the same domain. Specifically, we utilize the Natural Questions and TriviaQA datasets, each split into training and testing subsets. For the training questions, we generate query expansions and rank them according to retrieval effectiveness as detailed in the previous section. We then perform alignment using three different methods: Rejection Sampling Fine-Tuning (RSFT), Direct Preference Optimization (DPO), and a combination of RSFT followed by DPO.

The evaluation results are presented in Table~\ref{tab:natural} and Table~\ref{tab:trivia}, which show the top-N retrieval accuracy across different models.

On the Natural Questions dataset (Table~\ref{tab:natural}), zero-shot query expansion demonstrates mixed results, sometimes outperforming and sometimes underperforming the original query. The filtering approach shows improvement over the baseline, although RSFT alone lags behind filtering, DPO and the combination of RSFT and DPO outperform filtering, with RSFT + DPO providing the best results overall.

On the TriviaQA dataset (Table~\ref{tab:trivia}), zero-shot query expansion is generally less effective than the original query, and the filtering approach also underperforms compared to the zero-shot baseline. However, our alignment approaches, particularly DPO and RSFT + DPO, consistently achieve better results, with both methods performing nearly equally well at higher top-N accuracies. These results highlight the robustness of our alignment approach, as it consistently outperforms baseline methods across both the Natural Questions and TriviaQA datasets, demonstrating its effectiveness in various in-domain retrieval settings.

\subsection{Out-of-distribution Generalization}
To further evaluate the robustness and generalization capabilities of AQE, we conduct experiments where models are fine-tuned on one dataset and tested in a zero-shot setting on a different domain. This setup allows us to assess how well the models generalize beyond the training data.

Notably, the filtering approach struggles in this out-of-distribution setting, failing to surpass both the original query expansion and zero-shot query expansion, with a significant drop in top-1 retrieval accuracy by 5-10\%. This decline highlights the limitations of the filtering method in generalization to new domains.

In contrast, our alignment approach demonstrates more consistent performance across different domains, consistently outperforming the baselines. Among the tested methods, RSFT + DPO proves to be the most robust, yielding the highest retrieval accuracy across the board in out-of-distribution scenarios.

Although the filtering approach showed strong results in the original work by \citet{chuang2023expand}, its effectiveness is highly model-dependent and relies heavily on the specific generation-augmented retrieval (GAR) generators \citep{mao-etal-2021-generation} as noted by the authors. In contrast, our alignment methods are more model-agnostic, delivering strong performance across both in-domain and out-of-domain datasets, making them a more versatile solution for diverse retrieval tasks.

\begin{table*}[ht]
    \centering
        \caption{Top-N retrieval accuracy comparison of different query expansion methods, with filtering and alignment approaches trained on the Natural Questions and TriviaQA datasets, and tested on the WebQA dataset. Bold indicates the highest effectiveness for a given training set and $\ddagger$ superscript denotes statistically significant improvements over all the baselines (Original query, Query expansion, and Filtering) using Student's t-tests with Bonferroni correction.}
    \label{tab:webQA}
    \begin{tabular}{llllllll}
        \toprule
      & Model & Top-1 & Top-5 & Top-10 & Top-20 & Top-50 & Top-100 \\
        \midrule
        \midrule
       \multirow{2}{*}{Baselines} &  Original query & 18.9 & 41.8 & 52.1 & 62.4 & 71.6 & 75.5 \\
        & Query expansion (zero-shot) & 25.9 & 46.5 & 55.8 & 65.1 & 73.4 & 78.1 \\
        \midrule
        \multicolumn{7}{c}{Trained on Natural Questions} \\
        \midrule
        & Filtering & 9.1 & 22.9 & 29.7 & 36.5 & 47.7 & 54.8 \\ \hdashline \rule{0pt}{2.3ex}
       \multirow{3}{*}{AQE (Ours)} &  RSFT  & 28.0 & 49.2$^\ddagger$ & 59.4$^\ddagger$ & 67.1$^\ddagger$ & 74.7 & 79.0 \\
       &  DPO  & 31.2$^\ddagger$ & 51.5$^\ddagger$ & 61.3$^\ddagger$ & 67.9$^\ddagger$ & 75.0 & 78.3 \\
       &  RSFT + DPO  & \textbf{31.7}$^\ddagger$ & \textbf{53.4}$^\ddagger$ & \textbf{62.2}$^\ddagger$ & \textbf{70.4}$^\ddagger$ & \textbf{77.2}$^\ddagger$ & \textbf{80.9}$^\ddagger$ \\
        \midrule
        \multicolumn{7}{c}{Trained on TriviaQA} \\
        \midrule
       &  Filtering & 13.4 & 31.0 & 40.6 & 50.8 & 62.5 & 69.8 \\ \hdashline \rule{0pt}{2.3ex}
       \multirow{3}{*}{AQE (Ours)}  &  RSFT  & 28.7$^\ddagger$ & 48.1 & 56.9 & 64.5 & 73.4 & 78.5 \\
       &  DPO  & \textbf{30.7}$^\ddagger$ & \textbf{50.3}$^\ddagger$ & \textbf{59.0}$^\ddagger$ & \textbf{66.4} & \textbf{75.2} & \textbf{79.3} \\
       &  RSFT + DPO  & 27.7 & 48.8 & 57.7 & 65.8 & 73.9 & 79.1 \\
        \bottomrule
    \end{tabular}
    \vspace{10pt}
\end{table*}

\begin{table*}[ht]
    \centering
        \caption{Top-N retrieval accuracy comparison of different query expansion methods, with filtering and alignment approaches trained on the Natural Questions and TriviaQA datasets, and tested on the EntityQs dataset. Bold indicates the highest effectiveness for a given training set and $\ddagger$ superscript denotes statistically significant improvements over all the baselines (Original query, Query expansion, and Filtering) using Student's t-tests with Bonferroni correction.}
    \label{tab:EntityQs}
    \begin{tabular}{llllllll}
        \toprule
         & Model & Top-1 & Top-5 & Top-10 & Top-20 & Top-50 & Top-100 \\
        \midrule
        \midrule
        \multirow{2}{*}{Baselines}  & Original query & 44.2 & 62.3 & 68.9 & 73.7 & 78.3 & 81.2 \\
         & Query expansion (zero-shot) & 44.1 & 60.2 & 66.9 & 72.2 & 78.1 & 81.3 \\
        \midrule
        \multicolumn{7}{c}{Trained on Natural Questions} \\
        \midrule
        &  Filtering & 23.7 & 36.5 & 42.1 & 48.1 & 55.3 & 60.5 \\ \hdashline \rule{0pt}{2.3ex}
       \multirow{3}{*}{AQE (Ours)}  & RSFT  & 45.6 & 62.5 & \textbf{68.6} & 73.1 & \textbf{78.4 }& \textbf{82.0 }\\
     & DPO  & 43.6 & 59.4 & 65.2 & 70.0 & 76.6 & 79.7 \\
     & RSFT + DPO  & \textbf{46.7}$^\ddagger$  & \textbf{63.0} & \textbf{68.6} & \textbf{73.6} & 78.1 & 81.9 \\
        \midrule
        \multicolumn{7}{c}{Trained on TriviaQA} \\
        \midrule
        &  Filtering & 35.9 & 51.7 & 58.3 & 64.4 & 70.8 & 75.9 \\ \hdashline \rule{0pt}{2.3ex}
       \multirow{3}{*}{AQE (Ours)}  &  RSFT  & 41.8 & 58.0 & 64.9 & 70.5 & 76.6 & 81.2 \\
        &  DPO  & \textbf{47.0}$^\ddagger$ & \textbf{63.0} & \textbf{68.2} & \textbf{73.4} & \textbf{78.5} & {81.7} \\
        &  RSFT + DPO  & {45.1} & {61.2} & {67.9} & {72.8} & \textbf{78.5} & \textbf{82.0} \\
        \bottomrule
    \end{tabular}
    \vspace{10pt}
\end{table*}

\subsection{Computational Efficiency}
\begin{figure*}[t]
    \centering
    \includegraphics[width=\textwidth]{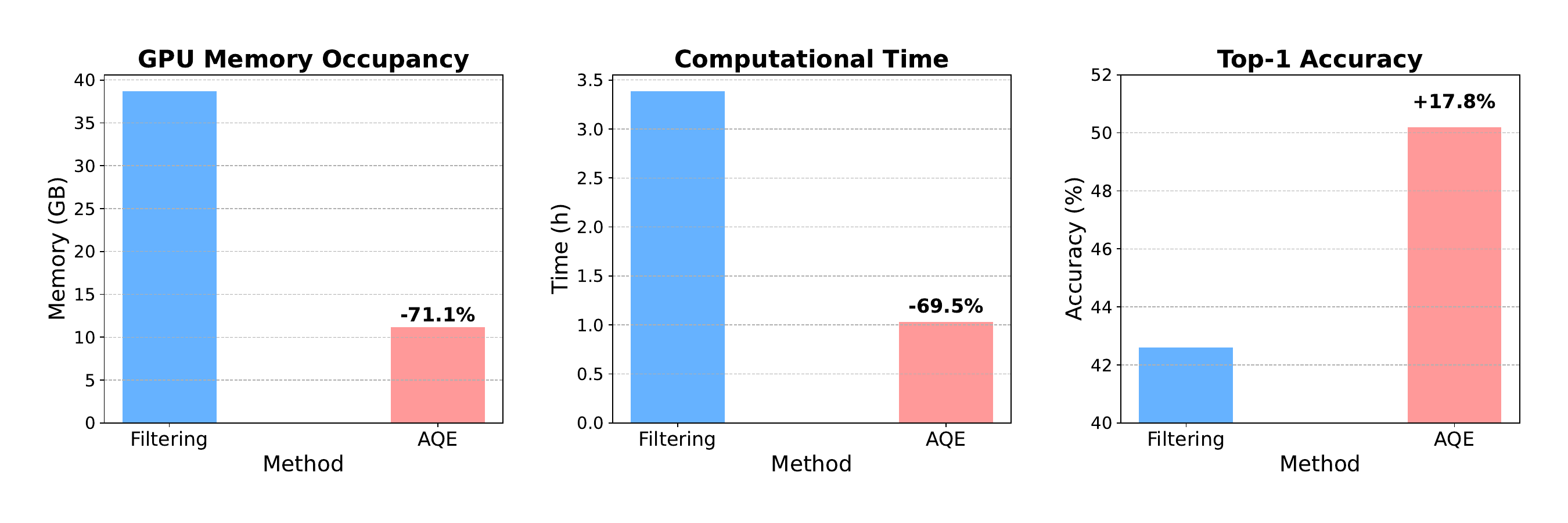}
    \caption{Comparison of GPU memeroy occupancy, computational time, and top-1 retrieval accuracy of filtering and AQE when performing inference on TriviaQA.}
    \label{fig:cost}
\end{figure*}

Latency is a key consideration for query expansion, as users must wait for the expansion and retrieval process to complete in real time. Figure~\ref{fig:cost} presents a comparison of computational memory and time between the filtering approach and AQE during inference on the TriviaQA dataset. AQE demonstrates significant improvements in efficiency, reducing memory consumption by 71.1\% and computational time by 69.5\%, while simultaneously enhancing top-1 retrieval accuracy by 17.8\%. These results highlight the effectiveness of AQE not only in improving retrieval performance but also in reducing the computational resources required, making it a much more scalable and practical solution for real-world applications.

\begin{figure}[t]
    \centering
    \includegraphics[width=0.45\textwidth]{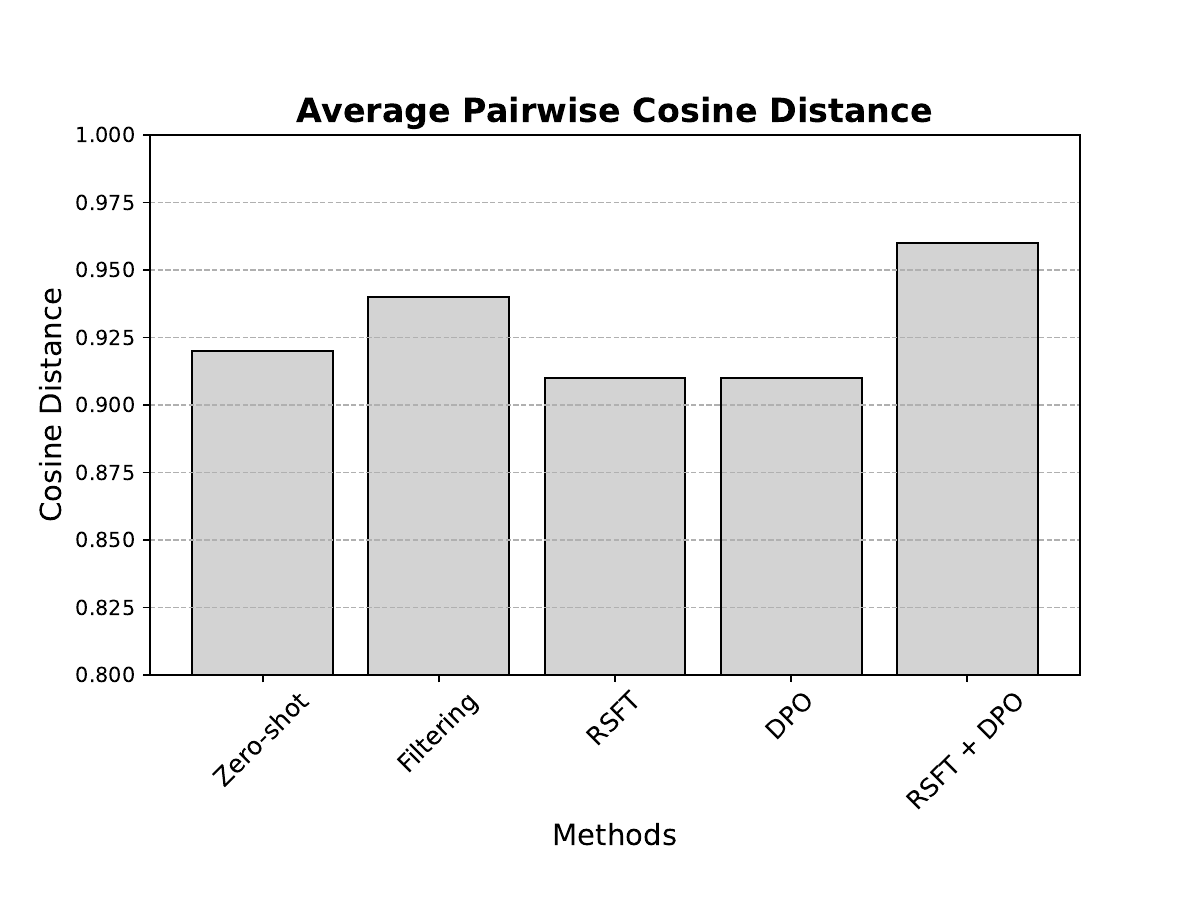}
    \caption{Diversity of the generated query expansions.}
    \label{fig:similarity}
\end{figure}

\subsection{Expansion Diversity Analysis}
To further investigate the differences in the query expansions generated by each method, we examine their \textit{generation diversity}. Specifically, we apply Sentence-BERT \citep{reimers-2019-sentence-bert} to obtain embeddings for each query expansion. We then compute the \textit{average pairwise cosine similarity} between each pair of embeddings to measure the diversity of the generated queries. Formally, for a set of query expansions \( Q = \{q_1, q_2, \dots, q_n\} \), where each \( q_i \) is a query expansion with an embedding \( \mathbf{e}_i \), we compute the diversity score as:
\[
D = \frac{2}{n(n-1)} \sum_{i=1}^{n-1} \sum_{j=i+1}^{n} \text{cosine\_similarity}(\mathbf{e}_i, \mathbf{e}_j)
\]
This provides a measure of how similar or diverse the generated expansions are—lower cosine similarity implies greater diversity.

The results are presented in Figure~\ref{fig:similarity}. While the filtering approach slightly improves generation diversity, RSFT and DPO individually tend to slightly reduce diversity, likely because these methods optimize for retrieval effectiveness. However, RSFT + DPO increases diversity. This improvement can be attributed to the combined effects of RSFT, which refines the model toward optimal expansions, and DPO, which explicitly optimizes for a balance between the best and worst expansions, encouraging the model to explore a broader range of queries while maintaining effectiveness.


\begin{table*}[h]
\centering
\caption{Combined query expansions from three methods on the Natural Questions dataset. Hint-based expansions are highlighted in \textcolor{hintcolor}{blue}, and direct answers are in \textcolor{answercolor}{red}.}
\label{table:combined_query_expansions}
\captionsetup{skip=8pt}
\begin{tabularx}{\textwidth}{X|X|X|X}
\hline
\textbf{Original Query} & \textbf{Filtering} & \textbf{RSFT} & \textbf{RSFT + DPO} \\
\hline\hline
who sings theme tune to orange is the new black &
\textcolor{hintcolor}{whose theme song is orange is the new black?} &
\textcolor{hintcolor}{the name of the singer} &
\textcolor{answercolor}{Regina Spektor sings the theme song} \\
\hline
who wrote lord have mercy on the working man &
\textcolor{hintcolor}{the writer} &
\textcolor{hintcolor}{the writer} &
\textcolor{answercolor}{Travis Tritt wrote the song} \\
\hline
how many books are in the diary of a whimpy kid series &
\textcolor{answercolor}{The series has 6 books.} &
\textcolor{hintcolor}{Count} &
\textcolor{answercolor}{There are 13 books} \\
\hline
what material was used to build the roofs of houses in burzahom &
\textcolor{answercolor}{Asphalt} &
\textcolor{hintcolor}{the material used to build the roofs in burzahom is wood} &
\textcolor{answercolor}{They used mud and reeds for roofing} \\
\hline
\end{tabularx}
\end{table*}




\subsection{Examples of Query Expansions}

Table~\ref{table:combined_query_expansions} presents examples of query expansions generated by the three methods. Hint-based expansions are highlighted in \textcolor{hintcolor}{blue}, while direct answers are in \textcolor{answercolor}{red}.

The Filtering method often generates a mix of hint-based expansions and partial answers. For instance, in the query \emph{``who sings theme tune to orange is the new black''}, Filtering returns a rephrased hint \textcolor{hintcolor}{``whose theme song is orange is the new black?''}, which does not directly answer the query but gives partial guidance. While Filtering sometimes generates partial answers, these tend to be less accurate than the direct answers provided by RSFT + DPO. For example, in the same query about the theme song, RSFT + DPO provides the accurate, direct answer \textcolor{answercolor}{``Regina Spektor sings the theme song''}, enhancing retrieval effectiveness. This trend is consistent across multiple examples, with RSFT + DPO generally offering more precise answers, leading to improved overall performance.

\section{Conclusion}
In this work, we introduced Aligned Query Expansion (AQE), a novel approach for enhancing query expansion in passage retrieval tasks. AQE leverages alignment techniques, such as Rejection Sampling Fine-Tuning (RSFT) and Direct Preference Optimization (DPO), to refine the query generation process, eliminating the need for extensive filtering methods. By aligning large language models to produce high-quality query expansions directly, AQE improves both retrieval effectiveness and computational efficiency.

Through comprehensive experiments, we demonstrated the effectiveness of AQE across multiple datasets. In in-distribution settings, AQE consistently outperformed baseline methods, including traditional query expansion approaches and filtering-based systems. 

Furthermore, AQE showed strong generalization capability in out-of-distribution scenarios, maintaining robust performance even when fine-tuned on one dataset and evaluated on another. 

Notably, AQE significantly reduced memory consumption and computational time while improving retrieval effectiveness. This makes AQE not only a powerful but also a scalable and practical solution for real-world retrieval systems, where latency and resource constraints are critical. By aligning large language models with RSFT and DPO, it eliminates costly filtering steps, improves retrieval performance, and reduces latency and memory usage—crucial for real-time, large-scale applications. Concurrent work proposing alignment of query expansion has been A/B tested and deployed, showing significant gains in a large industrial setting~\cite{peng2024large}.

Overall, the results of our experiments underscore the benefits of combining RSFT and DPO for generating diverse and relevant query expansions, resulting in a method that is both model-agnostic and adaptable to various retrieval contexts. Future work could explore further enhancements to the alignment strategies and apply AQE to more complex retrieval tasks in diverse domains.


\bibliographystyle{ACM-Reference-Format}
\bibliography{ref}

\appendix

\end{document}